# REACTION MECHANISMS IN PETROLEUM: FROM EXPERIMENTATION TO UPGRADING AND GEOLOGICAL CONDITIONS

F. Lannuzel [1], V. Burklé-Vitzthum [1], R. Bounaceur [1], P-M. Marquaire [1] and R. Michels [2]

[1] *DCPR CNRS-UMR 7630, ENSIC INPL Nancy University BP 20451, 54001 Nancy France*
[2] *G2R CNRS-UMR 7566, UHP Nancy University, BP 236, 54501 Vandœuvre lès Nancy France*

Abstract: Among the numerous questions that arise concerning the exploitation of petroleum from unconventional reservoirs, lie the questions of the composition of hydrocarbons present in deep seated HP-HT reservoirs or produced during *in-situ* upgrading steps of heavy oils and oil shales. Our research shows that experimental hydrocarbon cracking results obtained in the laboratory cannot be extrapolated to geological reservoir conditions in a simple manner. Our demonstration is based on two examples: 1) the role of the hydrocarbon mixture composition on reaction kinetics (the "mixing effect") and the effects of pressure (both in relationship to temperature and time). The extrapolation of experimental data to geological conditions requires investigation of the free-radical reaction mechanisms through a computed kinetic model. We propose a model that takes into account 52 reactants as of today, and which can be continuously improved by addition of new reactants as research proceeds. This model is complete and detailed enough to be simulated in large ranges of temperature (150-500°C) and pressures (1-1500 bar). It is thus adapted to predict the hydrocarbons evolution from upgrading conditions to geological reservoirs.

## 1. INTRODUCTION

The perspective of a shortening of petroleum reserves raises the challenge of developing new exploitation strategies concerning unconventional reservoirs, like for instance deep seated, high pressure-high temperature reservoirs, tar sands or oil shales. In deep seated reservoirs, the challenge concerns our capacity to be able to reconstruct properly the evolution of hydrocarbons composition during basin evolution in order to predict the fluid quality. Concerning heavy oils and oil shales rise for instance the questions of the composition of hydrocarbons after *in-situ* upgrading steps and the prediction of the formation of acid gases.

The proposed presentation deals with the difficulty to extrapolate experimental results outside of the range of conditions at which they were obtained. The examples of the role of the hydrocarbon fluid composition (the "mixing effects", i.e the kinetic effect induced by a hydrocarbon on a mixture) and of pressure on the kinetics of cracking are taken. We are presenting our methodology to answer some of these questions by combining experimentation, construction of reaction networks and kinetic modelling.

## 2. EXPERIMENTAL CONDITIONS AND CONSTRUCTION OF THE KINETIC MODEL

Reactants are chosen among the most representative compounds of the aliphatic and the aromatic fractions of a crude oil, typically normal and branched alkanes, naphtenes, toluene and other alkylaromatics, hydroaromatics and PAH. Reactants are however chosen and classified based on their kinetic properties (i.e. their capacity to modify the kinetics of cracking of the mixture) rather than on their structure. For instance, benzene, toluene and decylbenzene belong to the same structural family of the monoaromatics. Yet, the kinetic influence of each molecule on hydrocarbon mixtures cracking is fundamentally different: benzene has no kinetic effect, toluene can be a strong inhibitor (Razafinarivo, 2006), long chain alkylbenzenes are inhibitors (Burklé-Vitzthum *et al.*, 2004) but act differently than toluene. On this basis, we also consider that a specific compound can be representative of the behavior of several other molecules having the same kinetic effect. Thus, the behavior of n-octane in the hydrocarbon mixture is representative of all n-alkanes; toluene is representative of all methylaromatics and decylbenzene of all alkylaromatics bearing a side chain containing more that 4 carbon atoms. Of course, all of our assertings are verified by experimentation and chemical mechanisms construction (Burklé-Vitzthum *et al.*, 2003 and 2004; Lannuzel, 2007).

The model compounds are pyrolysed pure or in mixtures, typically at 350-450°C under 100-1000 bars. The experimental setting used is that of confined pyrolysis (Monthioux and Landais, 1988; Landais *et al.,* 1989). Reactants are loaded into gold cells under controlled neutral atmosphere. After sealing, the gold cells are loaded into stainless steel reactors and heated under controlled pressure. The choice of the reactor design is mainly argumented by its capacity to reproduce reaction results on kerogen, oils and model compounds compatible with those observed in sedimentary basins (Monthioux *et al.*, 1985; Landais *et al.*, 1994). After the experiments, a complete mass balance of the reaction products is made using thermodesorption - multidimensional gas chromatography (Gérard *et al.*, 1994). Compounds are identified using GC-MS.

After collection of the experimental data, the first step of reaction mechanisms writing is performed on the pyrolysis results of the pure compounds. Mechanisms are composed of elemental free-radical reactions with their corresponding rate constants that are not adjustable parameters. Second step concerns the binary mixtures, for which the mechanisms of both pure compounds are combined and the relevant cross reactions added. Other species can then be added incrementally to the mixture, adding their own reactions and cross reactions with the other species to the mechanism. The reaction model can thus be increased in size as research progresses.

When the model describes correctly the experiments, we consider that the most important reactions are very likely taken into account and the mechanism is thus validated. This model, based on the specific properties of elemental free-radical reactions can then be extrapolated to other pressure-temperature-time conditions, without empirical adjustment of rate constants by the user, as it is the case for simplistic kinetic models.

## 3. APPLICATION TO THE "MIXING EFFECTS"

In this section, we will show that "mixing effects" (co-reactions between compounds) can not be correctly extrapolated from experimental to reservoir conditions without a detailed kinetic model. We will compare the effect of an alkylaromatic with a long side-chain (decylbenzene) on n-alkanes cracking, to the effect of toluene. Then we will study the effect of toluene as a function of temperature and pressure.

### 3.1 Effects of decylbenzene on n-hexadecane cracking

An experimental study (Burklé-Vitzthum *et al.*, 2004) showed that n-hexadecane cracking rate is divided by a factor 2 when it is mixed with 20% of decylbenzene at 330°C and 700 bars. This effect has been explained by considering reaction mechanisms (1066 elemental reactions). The corresponding kinetic model allowed to extrapolate the results to the geological reservoirs time-temperature conditions (200°C-10 million years) and showed that decylbenzene indeed inhibits the n-hexadecane cracking, but by about a factor 50 to 3000, depending on temperature and on conversion. Inhibition can be measured by the Inhibition Factor (IF) that is defined by the ratio: conversion of pure alkane/conversion of alkane in mixture. If the IF = 1, no inhibition effect is observed. Figure 1 shows the evolution of the IF values vs temperature for the mixture hexadecane-decylbenzene (conversion = 1%). The kinetic role of long-chain aromatics at geological time-temperature conditions is thus much stronger than what could be inferred from the experimental conditions.

### 3.2 Effects of toluene on n-octane cracking at 350°C

Because of the similarity of structure between decylbenzene and toluene, we may think that they behave in a similar manner. At very low pressure (1mbar), toluene inhibits n-octane by about a factor 3 (Razafinarivo, 2006), but at 700 bar, this inhibition becomes negligible (about a factor 1.1; Lannuzel, 2007). So, at high pressure, toluene does not seem to be an interesting compound from a kinetic point of view.

The corresponding mechanism has been constructed (Lannuzel, 2007). It is composed of the mechanisms of pure toluene (30 elemental reactions), of pure octane and of the cross reactions (51 processes). In a general manner, the inhibition appears when a radical that is particularly stable is formed during the reaction. That is the case of the benzyl radical that is stabilized by resonance. But a kinetic analysis of the mechanism shows that the benzyl radical is consumed by addition on alkenes and so the inhibition effect does not appear at 350°C under 700 bar.

At very low pressures, the addition processes are negligible and so the accumulation of benzyl radicals in the medium causes the inhibition.

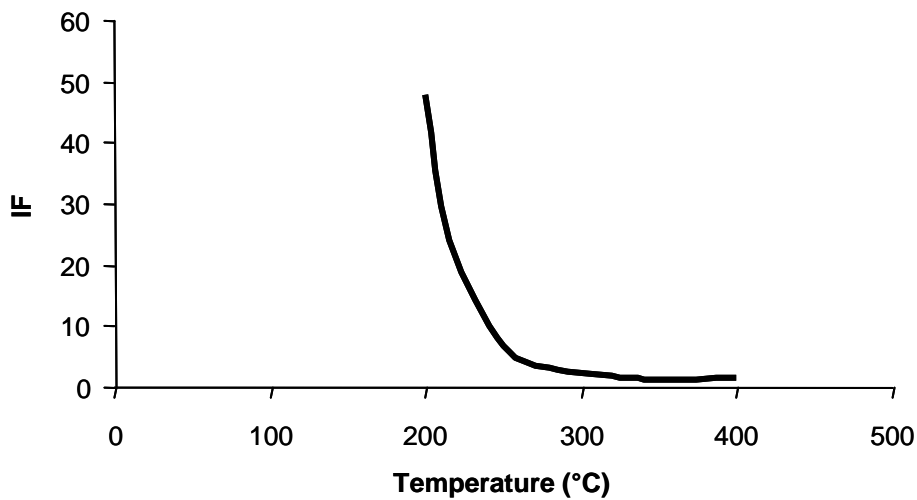

Fig. 1. Evolution of the Inhibition Factor as a function of the temperature for the mixture 80% n-hexadecane - 20% decylbenzene (conversion of hexadecane = 1%).

Concerning decylbenzene, it has been shown that the addition reactions of decylbenzene radicals on alkenes are less important (the rate constant is about 100 times lower than the corresponding reaction with the benzyl radical) because of steric reasons. So decylbenzene acts as a better inhibitor than toluene.

*3.3 Effects of toluene on n-octane cracking at low temperatures*

The mechanism of the mixture toluene-octane has been simulated at 200°C under various pressures. Under high pressure (and in the contrary to what has been observed in our experiments at 350°C), toluene acts as a good inhibitor on the n-alkanes cracking (Figure 2): the conversion of octane is divided by about 10-20, and its half life is more than 2 times greater.

At lower pressures, the inhibition is weaker. Figure 3 shows the variation of the IF values vs temperature (200-450°C) for several pressures. Without a detailed and robust mechanism, the variation of the IF value appears totally unpredictable.

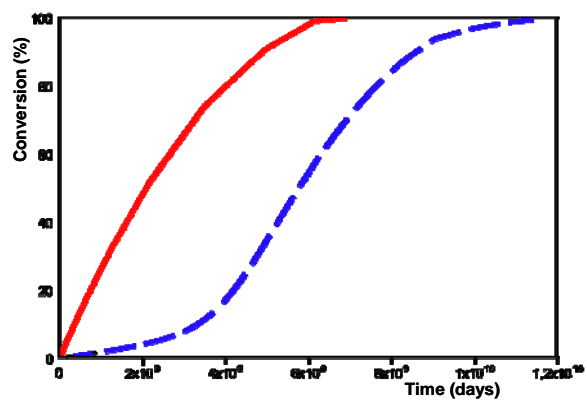

Fig.2. Conversion of octane, as a pure compound (continuous line) and in mixture with toluene (10 %) (dashed line) at 200°C, as a function of time.

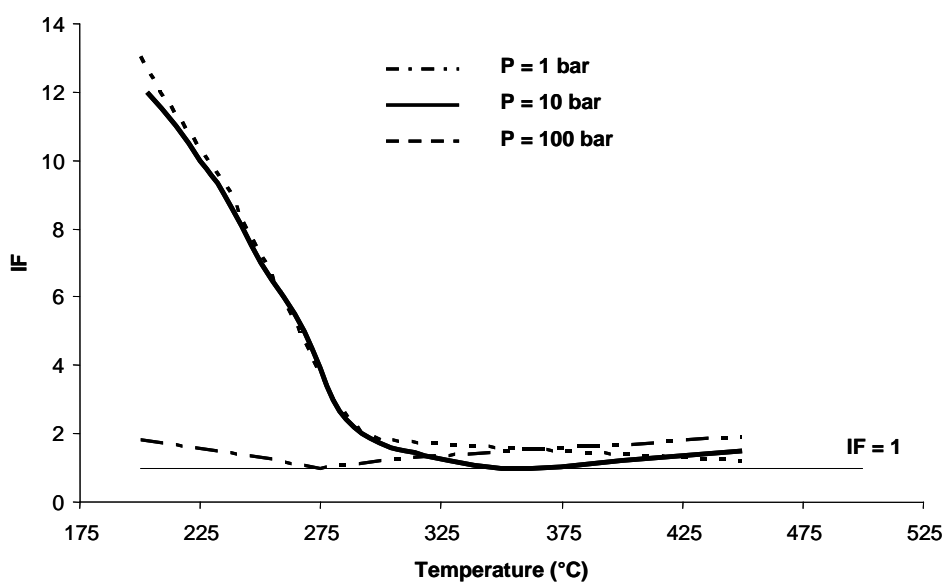

Fig.3. Evolution of the Inhibition Factor as a function of the temperature, for several pressures, in the case of the mixture 90% octane – 10% toluene (conversion of octane = 1 %).

The example of toluene has been chosen to demonstrate that the kinetic effect of a compound on the cracking of another reactant can not be easily predicted: the results of experiments can not be extrapolated outside the experimental temperature and pressure ranges without the mechanism. This conclusion is valid for secondary cracking in geological reservoirs as well as *in-situ* upgrading.

## 4. APPLICATION TO PRESSURE EFFECTS

Pressure is an important parameter to take into account, since its value changes considerably from deep seated petroleum reservoirs (several hundred to more than thousand bars) to *in-situ* upgrading of oil shales or heavy oils (few tens to more than hundred bars). Changing the pressure changes the concentrations of the compounds and so the importance of the bimolecular reactions compared to the unimolecular reactions. If the mechanism is detailed enough and takes into account all unimolecular and bimolecular processes, it is valid from very low pressures (1 bar) to very high pressures (about 1500 bar). Let us investigate the effect of pressure on the distribution of products and on the conversion obtained by alkanes cracking.

*4.1 Distribution of products*

Three types of products are obtained after n-alkane cracking: alkanes-minus (molecular weight less than the reactant), alkanes-plus (M.W greater than the reactant) and alkenes. The distribution between these three types is modified by temperature in a negligible manner, but pressure has a very strong effect (Figures 4). At low pressures, unimolecular processes, particularly decomposition by β-scission, are dominant; so alkenes are the most abundant products and no alkanes-plus are formed. At high pressures, bimolecular processes and particularly additions become important: no alkenes can be detected because they are consumed to form alkanes-plus.

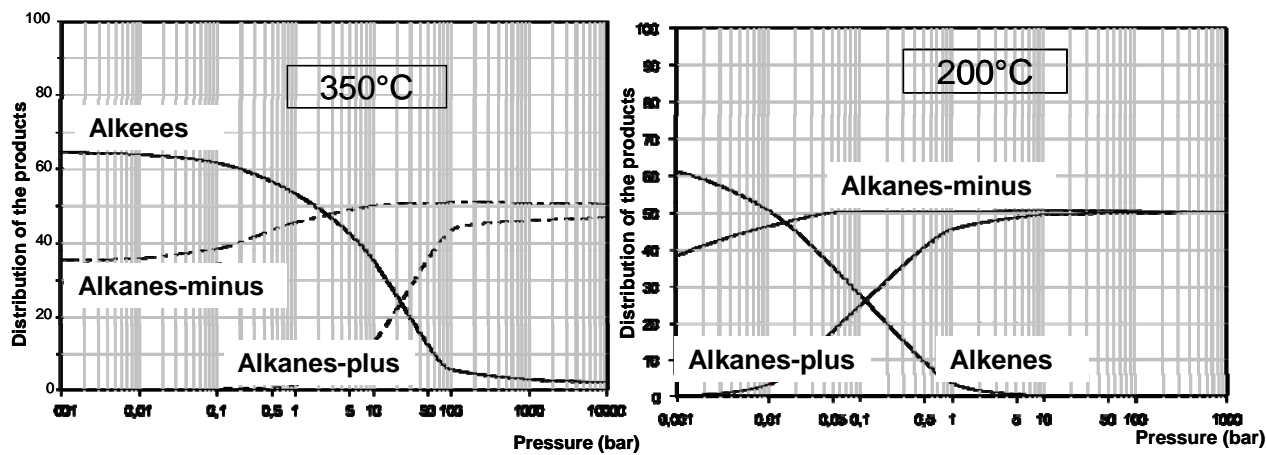

Fig.4. Distribution of the products obtained by a n-alkane cracking as a function of pressure, for two temperatures.

*4.2 Conversion*

The simulated evolution of the n-octane conversion vs pressure is presented Figure 5. A maximum is obtained whatever the temperature but, when the cracking temperature increases, a greater pressure is needed to obtain this maximum. So at very low and very high pressures, the n-octane cracking seems to be inhibited by pressure. A precise analysis of the mechanism reveals that the predominant free-radicals are different at low and high pressures and so the predominant processes are different too. The maximum occurs when no specific type of radicals is predominant but when all of them are present.

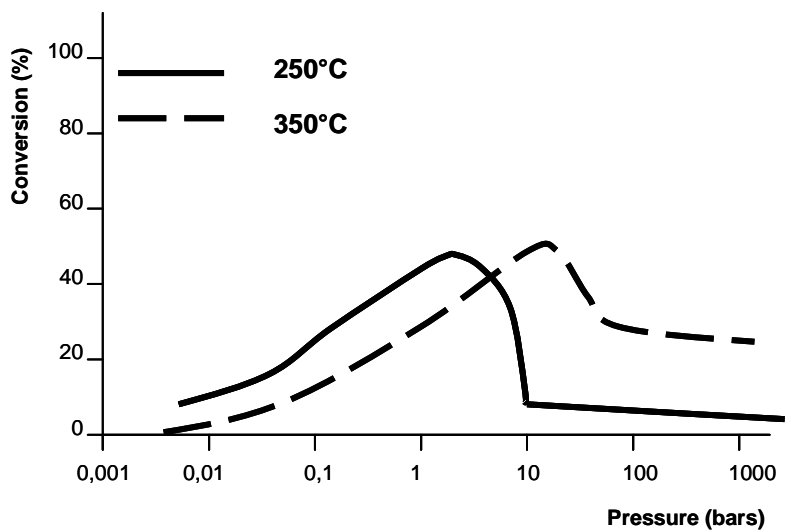

Fig.5. Conversion of a n-alkane as a function of pressure for two temperatures.

The understanding of the effects of pressure on the reaction mechanisms of a given hydrocarbon mixture composition can thus be evaluated by our kinetic model considering geological reservoir conditions. While no pressure effect is observed at 350°C, the influence of pressure could be quantitated at 200°C. This is, to our knowledge, the first quantitative modelling of the effects of pressure on the cracking at geological conditions.

## 5. CONCLUSION

Through the examples of the mixing effects (co-reactions) and of the pressure effects on hydrocarbon cracking, we have shown that any extrapolation of results outside of the range of experimental conditions can lead to severe errors when detailed kinetic model based on reaction mechanisms are not considered. Our model comprises 52 reactants for the moment (Dominé *et al.*, 2002) and it has been validated by correctly predicting the stability of some HP-HT reservoirs. Its construction principle permits to add more reactants to the model in a rigorous and relatively simple manner, thus allowing to investigate increasingly complex fluids. Such a model can represent correctly the aliphatic and the aromatic fractions of most crude oils, but the polar fractions are not explicitly taken into account. Nevertheless the asphaltenes can be treated as sources of hydrocarbons that constantly feed the model with reactants.

Our model is complete and detailed enough to be simulated in large ranges of temperature (150-500°C) and pressure (1-1500 bar). So it can be useful to predict the hydrocarbons evolution in various conditions: 1) secondary cracking of crude oils in geological reservoirs, 2) *in-situ* and *ex-situ* upgrading during the assisted recuperation of heavy oils and oil shales by heating or 3) by addition of gases or solvents. The next steps of this study will focus on $H_2S$ and hydrocarbons containing sulphur. Indeed one of the challenges of the assisted recuperation is to predict the formation of $H_2S$ and its effect on the upgrading of oil. Among the questions that arise: How is $H_2S$ formed? $H_2S$ leads to a stabilised free-radical; so, does $H_2S$ limit the upgrading of oil? How does it react with hydrocarbons? Most of the C-S bonds are relatively weak; are oils rich in sulphur more sensitive to cracking? Can we quantitate the induction effects due to C-S and S-S bonds breakage?

Acknowledgment: The authors would like to thank Total for funding this work, and especially D. Dessort for his interest and support.